\documentclass[a4paper,12pt]{article}
\usepackage{graphicx}
\title{Barometric pumping effect for radon-due neutron flux in underground laboratories}

\author{Yu V Stenkin$^{1,2}$, V V Alekseenko$^1$, D M Gromushkin$^2$,\\ O B Shchegolev$^1$ and V P Sulakov$^3$ }
\begin{document}
\maketitle
\footnotesize{%
$^1$ Institute for Nuclear Research of Russian Academy of Sciences, Moscow 117312\\
$^2$ National Research Nuclear University MEPhI, Moscow 115409\\
$^3$ Skobeltsyn Institute for Nuclear Physics, Moscow State University, Moscow 119991}\\

\thanks{  Correspondig author Yu. Stenkin: stenkin@sci.lebedev.ru}

\abstract{It is known that neutron background is a big problem for low-background experiments in underground Laboratories. Our global net of en-detectors sensitive to thermal neutrons includes the detectors running both on the surface and at different depths underground. We present here results obtained with the en-detector of 0.75 m$^2$ which is running more than 3 years in underground room at a depth of 25 m of water equivalent in Skobeltsyn Institute of Nuclear Physics, Moscow. Spontaneous increases in thermal neutron flux up to a factor of 3 were observed in delayed anti-correlation with barometric pressure. The phenomenon can be explained by a radon barometric pumping effect resulting in similar effect in neutron flux produced in $(\alpha,n)$-reactions by alpha-decays of radon and its daughters in surrounding rock.    
}	

\section{Introduction}
   Many low-background experiments, such as double-beta decay or dark matter searches, carried deep underground, are sensitive to a background caused by neutrons. Special passive and$/$or active protection against this background can not exclude it in full. Therefore, the experimentalists have to know neutron background variations and take it into account. As we have found, spontaneous increases of thermal neutron flux underground can be explained by a geophysical effect known as $\emph {barometric pumping effect}$  $\cite{1,2,3}$ for underground gases. When atmospheric pressure falls down it causes soil air and other soil gases (including radon) diffusion from deeper soil layers to higher layers and from soil to atmosphere. If underground laboratory has no special interior covering isolating it from soil gases, has connection with barometric air pressure and has no forced ventilation then radon inside its volume is following the barometric pumping effect. Moreover, radon concentration in surrounding rock or soil as well as radon-due neutron flux are affected by this effect. 
   In this work we show for the first time the barometric pumping effect observed in thermal neutron flux underground. 

\section{Detector}
A novel type of thermal neutron scintillation detector (${\it en-detector}$) is used. Solid and well known specialized granulated alloy scintillator $^6$LiF+ZnS(Ag) is used to detect thermal and epithermal neutrons. Thickness of the scintillator layer is $\sim$30 mg$/cm^2$. Sensitive area of scintillator surface is ~0.75 m$^2$. The detector recording efficiency for thermal neutron is equal to $\sim$20$\%$. The efficiency follows usual for thin detectors ~1/v dependence on neutron velocity v. Due to thin scintillator layer relativistic charged particles, such as electrons or muons, produce very small signals lying below the FADC threshold. This allows us to use the detector in scalar mode to study thermal neutron low flux variations. Pulse shape originated from heavy particles ($\alpha$ and triton) produced by neutron capture on $^6$Li target differ from that produced by relativistic charged particles or by PMT’s noise.
A distinctive  feature of our data acquisition process is that all pulses from PMT are amplified by a charge sensitive preamplifier and then digitized by a FADC or, in other words, preliminary full pulse shape analysis and selection are used to count real neutron pulses and to reject (but still counting them) the noise pulses. Our technique details can be found elsewhere $\cite{4}$. The en-detector is located in a mine at a depth of 25 m of water equivalent in a shaft of muon detector of EAS-MSU array (Moscow State University). Underground room where the detector is running for $\sim3$ yeas has no forced ventilation and has simple concrete walls and ceiling. The data are accumulated for a long period as a time series with 1 min step. The detector has very stable long-term characteristics which are controlled by daily energy deposit spectra. Environmental parameters in the room (air pressure P, temperature t and relative humidity h) are continuously measured using special digitized sensors.  

\section{Results and analysis}

The detector location is deep enough ($\sim$25 hadron interaction lengths) to prevent cosmic ray hadrons penetration to the detector vicinity. At this condition the detector records mostly the neutrons produced in ($\alpha$,n)-reactions (A,Z)+ $\alpha$ = (A+3, Z+2) + n on light crust elements (such as Be, B, F, Na, Al, Mg, Si) induced by natural $^{238}$U and $^{232}$Th alpha-radioactive chains nuclides $\cite{5}$. Therefore one could expect an absence of barometric effect at high depth. We tried to find the barometric coefficient using standard procedure but it was found to be close to 0. Nevertheless, sometimes we have observed spontaneous counting rate increase up to a factor of 3 as it is shown in fig. 1. Upper panel shows barometric pressure behavior which can explain our result. Long duration decrease of air pressure from Jan. 4 to Jan. 11, 2014 caused significant increase of the detector counting rate. (Note in brackets that increase of barometric pressure does not change the counting rate significantly).  
\begin{figure}[!htb]
	\includegraphics[width=12cm]{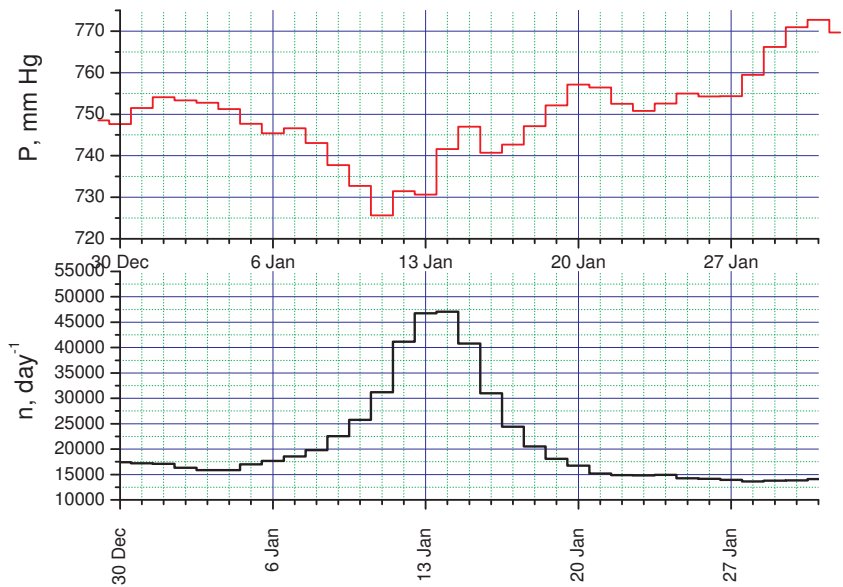}\hspace{4pc}%
	\caption{An example of air pressure and thermal neutron flux behavior (1-day points).}
	\end{figure}
    Radioactive gas Rn-222 originated from U-238 chain has 3.8-day half-life time and as an inert gas can migrate with air for a long distance in porous media including soil, sedimentary rock, concrete, etc. If radon or its daughter nuclide decays inside the above materials it can produce neutrons which can be thermalized there and then escape to the room volume. Decays in air (excluding the nearest vicinity of solid matter) are not accompanied with neutron productions due to absence of needed target in air and because pass length of several MeV alphas in air is equal only about 3-5 cm. This is why our detector records neutrons produced in surrounding rock or concrete ($\sim$2-3 m thickness) and thus is not sensitive to air ventilation and drought. The latter gives en-detector an advantage in comparison with ordinary radon-meters, which measure only concentration of radon in air and are sensitive to any air movement resulting in instability.   
    Looking again at fig.1 one can see that there exists a 2-day delay between the minimum in pressure and maximum in neutrons. To check it we made an analysis of the data trying to find an optimal delay between pressure and neutrons counting rate. Result is shown in fig. 2. It is seen that optimal shift is close to 2 days (-2.0$\pm0.1$ d) and correlation is rather good in optimal point.  
  \begin{figure}[htb!]
  	\includegraphics[width=12cm]{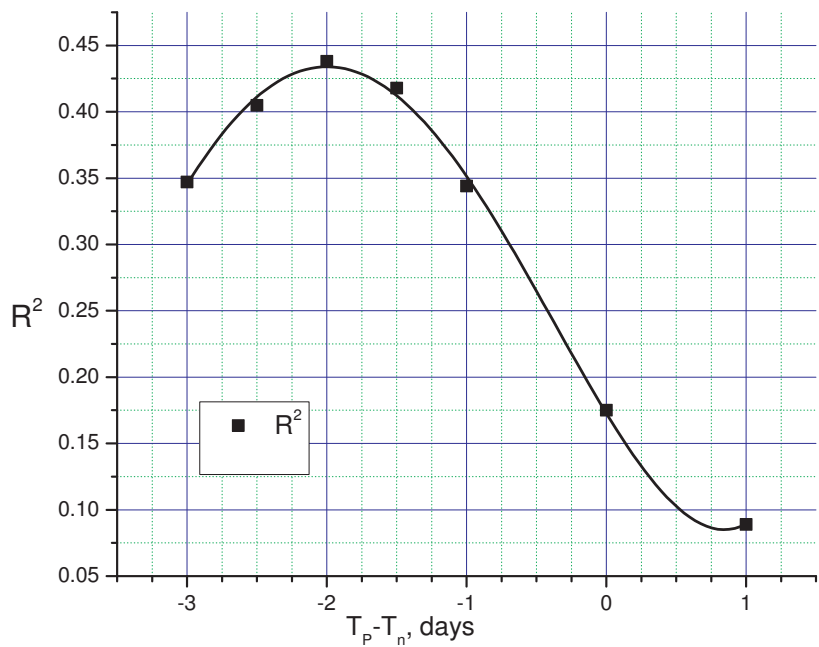}\hspace{2pc}%
  	\caption{\label{label} Square of correlation coefficient R as a function of time shift in data. }
  \end{figure}
    
This confirms that barometric pumping effect does exist not only for radon but for thermal neutrons as well. Applying this delay to pressure data we checked again the correlation between change in pressure $\Delta P$ and normalized neutron counts $\Delta n /<n>$ change. The result of 2 years data analysis is shown in fig. 3 where rather strong anti-correlation is observed. It is noticeable that the obtained delayed negative barometric coefficient $\beta$ is by a factor of $\approx 5.5$ higher of that for cosmic ray hadrons at the Earth surface being equal to $\sim1\%/$mm Hg.
  \begin{figure}[htb!]
  	\includegraphics[width=13cm]{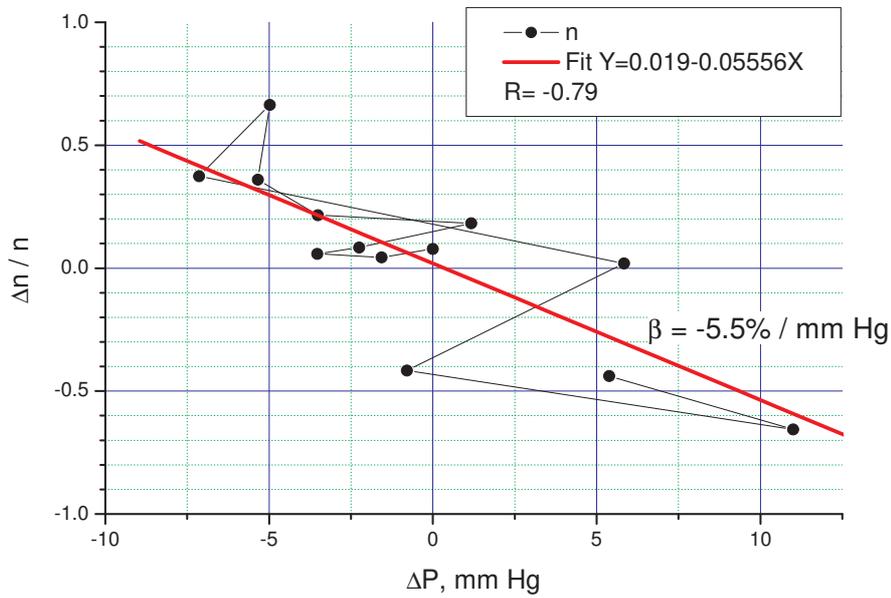}\hspace{2pc}%
  	\caption{\label{label} . Anti-correlation plot between 2-days delayed air pressure change and normalized neutron counting rate change.}
  \end{figure}

\section{Discussion }

To explain the observed delay one could imply the diffusion theory to gases in porous media. In simplified one dimensional approach one can write gas concentration C depending on time t and distance x as:\begin{equation}
C(t,x)=C_{0} \times exp(-x^{2}/(2Dt))
\end{equation} 
\\ where $C_{0}$ being initial concentration and D being diffusion coefficient. 
Integrating it over x we obtain:
 \begin{equation}
Q(t)=\int_0^\infty C(t,x)dx=\sim{\sqrt{tD}}
\end{equation} 
i. e. if constant pressure gradient is applied for a long time then Q(t) $\sim\sqrt{t}$. This result is similar to that 
shown in $\cite{6}$.
Taking into account radioactive decay of radon with lifetime equal to $\tau$, resulting time dependence looks like 
\begin{equation}
W(t) \sim Q(t)\times exp(-t/\tau)\sim\sqrt{tD}\times exp(-t/\tau). 
\end{equation} 
This function has a maximum at time $t_{max} =\tau /2 \approx 2.7 d$. This estimation is close (but not equal) to the observed 2-days delay between stopping of diffusion process (air pressure decrease stop) and visible maximum in neutron counts increase position. When pressure gradient change sign then accumulated radon in rock follows radioactive decay with above $\tau$ thus resulting in decrease of counts (from Jan. 14 to Jan. 20 in fig. 1) with $\tau \approx 5.5 d$. Note that the above estimations are qualitative and correct calculations have to be done by geophysicists.  

\section{Summary}

1. The effect of barometric pumping has been observed for the first time in thermal neutron flux underground.\\
2. Due to this effect a spontaneous increase of neutron background in underground laboratories which do not keep air pressure stable and have no hermetic shield against outside gases, could be as much as by a factor of 3.\\ 
3. Low background Labs should be careful when estimating neutron background capable to mimic the effect and have to control both neutron flux and air pressure inside the experimental volume underground. Up to date we did not find any underground physics publications (including the most recent one $\cite{7}$) mentioning this problem. \\
4. The observed effect is much higher than seasonal thermal neutron wave as well as tidal waves reported by us before at TAUP-2009 $\cite{8}$.

\section{Acknowledgments}
Authors acknowledgment financial support from RFBR grants $\#$14-02-00996 and $\#$13-02-00574. Authors are grateful to SINP MSU for providing us with an underground room facility.

\end{document}